\input ./style/arxiv-general.cfg
\documentclass[aoas,MSNbibl,nameyear,seceqn,rotating,dvips]{arximspdf}
\makeatletter
   \@ifpackageloaded{graphicx}{}{\usepackage{graphicx}}
\makeatother
\usepackage{mathbh,dcolumn}

\doi{10.1214/15-AOAS841}
\volume{9}
\issue{3}
\pubyear{2015}
\firstpage{1226}
\lastpage{1246}
\docsubty{FLA}

\makeatletter
\newcolumntype{d}[1]{D{.}{.}{#1}}
\makeatother

\begin{document}
\begin{frontmatter}

\title{Quantile regression for mixed models with an application
to examine blood pressure trends in China\thanksref{T1}}
\runtitle{Quantile regression for mixed models}

\begin{aug}
\author[A]{\fnms{Luke B.}~\snm{Smith}\corref{}\ead
[label=e1]{lukebrawleysmith@gmail.com}},
\author[A]{\fnms{Montserrat}~\snm{Fuentes}\ead[label=e2]{fuentes@ncsu.edu}},
\author[A]{\fnms{Penny}~\snm{Gordon-Larsen}\thanksref{T2}\ead[label=e3]{pglarsen@unc.edu}}
\and
\author[A]{\fnms{Brian J.}~\snm{Reich}\ead[label=e4]{bjreich@ncsu.edu}}
\runauthor{Smith, Fuentes, Gordon-Larsen and Reich}
\affiliation{North Carolina State University and University of
North Carolina at Chapel Hill}
\address[A]{Department of Statistics\\
North Carolina State University\\
Box 8203/5109 SAS Hall\\
Raleigh, North Carolina 26795-8203\\
USA\\
Department of Nutrition and Carolina Population Center\\
Gillings School of Public Health \& School of Medicine\\
University of North Carolina at Chapel Hill\\
137 E. Franklin Street\\
Chapel Hill, North Carolina 27516\\
USA\\
\printead{e1}\\
\phantom{E-mail: }\printead*{e2}\\
\phantom{E-mail: }\printead*{e3}\\
\phantom{E-mail: }\printead*{e4}}
\end{aug}
\thankstext{T1}{Supported in part by NSF Grant DMS-11-07046, NIH Grant
SR01ES014843, as well as NHLBI (R01-HL108427).}
\thankstext{T2}{The China Health and Nutrition Survey is funded by
NICHD (R01-HD30880),
(although no direct support was received from the grant for this
analysis), and supported by the Carolina Population Center (R24 HD050924).}

%
\received{\smonth{9} \syear{2014}}
%
\revised{\smonth{5} \syear{2015}}

%
\begin{abstract}
Cardiometabolic diseases have substantially increased in China in the
past 20 years and blood pressure is a primary modifiable risk factor.
Using data from the China Health and Nutrition Survey, we examine blood
pressure trends in China from 1991 to 2009, with a concentration on age
cohorts and urbanicity.
Very large values of blood pressure are of interest, so we model the
conditional quantile functions of systolic and diastolic blood pressure.
This allows the covariate effects in the middle of the distribution to
vary from those in the upper tail, the focal point of our analysis.
We join the distributions of systolic and diastolic blood pressure
using a copula, which permits the relationships between the covariates
and the two responses to share information and enables probabilistic
statements about systolic and diastolic blood pressure jointly.
Our copula maintains the marginal distributions of the group quantile
effects while accounting for within-subject dependence, enabling
inference at the population and subject levels.
Our population-level regression effects change across quantile level,
year and blood pressure type, providing a rich environment for inference.
To our knowledge, this is the first quantile function model to
explicitly model within-subject autocorrelation and is the first
quantile function approach that simultaneously models multivariate
conditional response.
We find that the association between high blood pressure and living in
an urban area has evolved from positive to negative, with the strongest
changes occurring in the upper tail.
The increase in urbanization over the last twenty years coupled with
the transition from the positive association between urbanization and
blood pressure in earlier years to a more uniform association with
urbanization suggests increasing blood pressure over time throughout
China, even in less urbanized areas.
Our methods are available in the R package \textbf{BSquare}.
\end{abstract}

%
\end{frontmatter}

\section{Introduction}\label{intro}
Globally, cardiovascular disease accounts for approximately 17 million
deaths a year, and nearly one third of the total causes of death in
2008 [\citet{WHO2011}].
Of these, complications of hypertension account for 9.4 million deaths
worldwide every year [\citet{Lim2013}].
Maximum (systolic) blood pressure and minimum (diastolic) blood
pressure are physiologically correlated outcomes but are differentially
affected by environmental factors [\citet{Benetos2001,Chobanian2003,Choh2011,Egan2010,Franklin2009,Luepker2012,Sesso2000}].
Most studies construct a combined measure using hypertension cutpoints
rather than looking across the distribution.
Systolic blood pressure (SBP) and diastolic blood pressure (DBP) have
differential effects on cardiovascular disease events [\citet
{Benetos2001,Franklin2009,Sesso2000,Stokes1989}],
so we model the conditional quantile functions of SBP and DBP.
In linear quantile regression the quantiles of the response (e.g., the
90th percentile) change linearly with the predictors, and the
regression effects are contingent upon the percentile chosen.
This allows the effects of urbanization to change along the
distribution of blood pressure, enabling sharper insight into the
relationship between urbanization and hypertension.
By conducting inference in the upper tails of SBP and DBP, we are able
to examine how urbanization affects individuals most at risk for hypertension.
Quantile regression also allows comparisons of conditional 90th
percentiles at different levels of a covariate.
We use longitudinal data from the China Health and Nutrition Survey
[\citet{Popkin2010}] to study the impact of urbanicity on those trends.
China provides an outstanding case study given recent and rapid
modernization and substantial concomitant environmental change.

In developed countries that experienced slow rates of modernization,
studies have suggested declines or leveling off in mean blood pressure
over the last century, potentially due to increased hypertension
treatment [\citet{Burt1995,Luepker2012,Egan2010,Mccarron2001}].
However, in China during a period of rapid modernization, we observed a
substantial increase in mean SBP and DBP over time, particularly in low
urbanicity areas [\citet{Attard2015}].
Understanding the association with urbanization across the full
distribution of blood pressure will allow researchers and policymakers
to understand the points along the distribution that may be most
amenable to environmental change, allowing more tailored intervention
targeting in China and in other low to middle income countries
undergoing similar urbanization.

The purpose of our paper is to address two important gaps in the
hypertension literature: (1) lack of attention to continuous SBP and
DBP as correlated outcomes; and (2) attention to the tails of their
distribution to examine how an environment-related exposure, in this
case urbanization, is associated with the distribution of SBP and DBP.
We utilize quantile regression to permit tail inference of SBP and DBP.
However, the China Health and Nutrition Survey (CHNS) data are
longitudinal in nature and we are interested in a multivariate outcome,
and current quantile regression methods would not permit satisfactory
exploration of our scientific aims.

Several previous approaches in the longitudinal literature simply
ignore the within-subject dependence when estimating the marginal
quantile effects.
\citet{Wang2011} constructed an empirical likelihood under the GEE
framework, then adjusted for the dependence in the confidence intervals.
For censored data, \citet{Wang2009} ignored the within-subject
dependence when estimating the marginal effects and controlled for the
within-subject dependence when conducting inference via a rank score test.
While these estimators are consistent, ignoring the within-subject
dependence for estimation can result in a loss of efficiency and undercoverage.
Another avenue is to introduce dependence via random intercepts, as in
\citet{Koenker2004}.
\citet{Waldmann2013} and \citet{Yue2011} assumed asymmetric Laplace
errors and included a random subject effect in the location parameter.
Presenting separate methodology for marginal and conditional inference,
\citet{RBW2010} accounted for within-cluster dependence via random
intercepts and flexibly modeled the density using an infinite mixture
of normals.
\citet{Jung1996} preserved marginal effects by incorporating correlated
errors in a quasi-likelihood model.
These models account for within-subject dependence via a location
adjustment for each cluster, which may not be sufficiently flexible.
Models that incorporate random slopes include \citet{Geraci2013}, who
used numerical integration to average out random effects for marginal inference,
and the empirical likelihood of \citet{Kim2011}.
The marginal effects of \citet{Geraci2013} do not necessarily maintain
their original interpretation after integrating over the random effects.
\citet{Kim2011} permit subject-specific inference for clustered data.
While these methods account for dependence, we are interested in
inference at the population level of temporally correlated data.
\citet{Jung1996} incorporates temporally correlated errors within a
subject, at the cost of assuming the response is distributed Gaussian.
Collectively these models lack attributes needed for our research
question, which relates to understanding the effects of urbanization on
SBP and DBP as correlated outcomes.
First, we want to conduct inference at multiple quantile levels without
assuming our response is distributed Gaussian.
The approaches above model one quantile level at a time and can result
in ``crossing quantiles'' [\citet{Bondell2010}], where for certain
values of the predictors the quantile function is decreasing in
quantile level.
Second, we need to model the autocorrelation within a subject to
maintain nominal coverage probabilities.
Third, for our application we anticipate that the effect of a covariate
on SBP may be similar to its effect on DBP, so we want a bivariate
model to facilitate communication across blood pressure type.

In this paper we introduce a mixed modeling framework for quantile
regression with these necessary attributes.
We accomplish these methodological innovations by extending the model
of \citet{RS2013} to accommodate autocorrelation and multiple responses.
In the random component we account for the dependence across time and
response via a copula [\citet{Nelsen1999}].
This permits the relationships between the covariates and the two
responses to share information and enables probabilistic statements
about SBP and DBP jointly.
Our copula approach maintains the marginal distributions of the
population-level quantile effects while accounting for within-subject
dependence, enabling inference at the population and subject levels.
Copulas previously utilized in the longitudinal literature [\citet
{Smith2010,Sun2008}] focused on mean inference and do not account for
predictors.
\citet{Chen2009} use a copula to account for serial dependence in
quantile estimation, without predictors.
Copulas have a straightforward connection to quantile function
modeling, as both rely on connecting the response to a latent uniformly
distributed random variable.
Our copula model resembles the usual mixed model [\citet{Diggle2002}] in
that covariates affect both the marginal population distribution via
fixed effects and subject-specific distributions via random slopes.
In the fixed component we allow for different predictor effects across
quantile level, response and year.
Our model is centered on the usual Gaussian mixed model, and contains
it as a special case.

We present a multilevel framework that extends the current Gaussian
mixed model to the quantile regression domain.
Our model permits examination of how urbanization has affected the
distributional tails of SBP and DBP, while controlling for the
dependence within CHNS subjects.
This allows us to draw new inferences in a more flexible manner than
mixed models where only the mean is affected by covariates.
For example, we can examine how regression effects in the lower tail,
middle of the distribution and upper tail change over time, and we can
examine how the quantiles of multiple responses adapt to changes in a predictor.
To our knowledge, this is the first quantile function model for
temporally-correlated responses within a subject and the first quantile
function model that accommodates a multivariate response with covariates.
We describe the data in Section~\ref{Data}.
In Section~\ref{Methods} we describe the mixed effect quantile model in
the univariate and multivariate cases.
In Section~\ref{Sim} we show the results of a simulation study that
illustrates the need to account for within-subject dependence in a
quantile framework.
In Section~\ref{Analysis} we analyze hypertension and we conclude in
Section~\ref{Conclusions}.

\section{Data}
\label{Data}
The CHNS was designed in 1986 to gauge a range of economic,
sociological, demographic and health questions [\citet{Popkin2010}].
The CHNS is a large-scale household-based survey drawing from 228
communities which were cluster sampled from 9 provinces.
Community structures include villages, townships, urban neighborhoods
and suburban neighborhoods.
The communities sampled are designed to be economically and
demographically representative of China.
Procedures for collecting the data are described in \citet{Adair2014}.
We use data collected in 7 waves, starting in 1991 and ending in 2009.
We focus on the Shandong province, located in central China, where
hypertension rates are elevated [\citet{Batis2013}].
We utilize the urbanicity index of \citet{Jones2010}.
Rather than dichotomizing communities into urban/rural groups, for each
wave \citet{Jones2010} assigned 0--10 scores for each of 12 factors,
including population density, economic activity, traditional markets,
modern markets, transportation, infrastructure, sanitation,
communications, housing, education, diversity, health infrastructure
and social services.
\citet{Jones2010} used factor analysis to confirm these factors
represent one latent construct.

%
\begin{figure}

\includegraphics{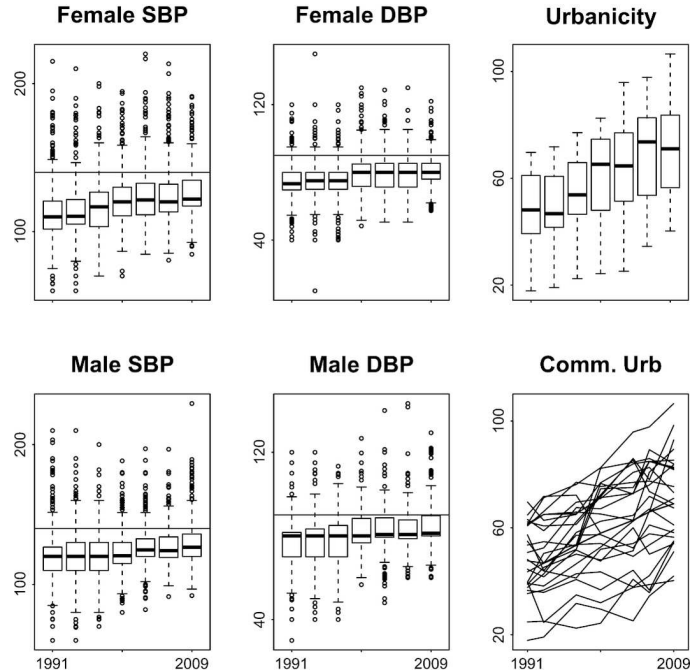}

\caption{Systolic blood pressure (SBP) and diastolic blood pressure
(DBP) by gender and urbanicity scores across time.
Blood pressure measurements are in millimeters of mercury (mmHg).
Urbanicity is a composite score measuring 12 features of the community
environment [\citet{Jones2010}].
Horizontal lines represent thresholds for high blood pressure, located
at 140~mmHg and 90~mmHg for systolic and diastolic blood pressure, respectively.}\label{f:exploratory}
\end{figure}

We have two scientific goals for these data.
First, we want to estimate the role of urbanicity in these trends.
Second, we want to examine blood pressure trends over time across
different age cohorts.
We bin individuals into six age groups: age $< 18$, $18 \leq$ age $<
30$, $30 \leq$ age $< 40$, $40 \leq$ age $< 50$, $50 \leq$ age $< 60$
and age $\geq60$.
For individuals with age $< 18$, blood pressure is very correlated with
height and age, rendering uninterpretable comparisons across children
and adults.
For this reason, most studies focus on children or adults, and we focus
on adults in this paper.
The plots in Figure~\ref{f:exploratory} show slight increases over time
in blood pressure across both genders and large increases over time in
urbanicity.
We construct an urbanicity by age interaction effect to look for
associations between urbanicity and different age cohorts.
As in \citet{Attard2015}, we stratify our analyses by gender.
Other covariates include current smoking status (men only due to low
female rates) and current pregnancy status (female only).
To look for changes across time, we include temporal linear trends for
all predictors.

\section{Methods}
\label{Methods}
In this section we present our methods for mixed model quantile regression.
We first specify the marginal quantile functions in Section~\ref{Marginal}.
In Sections~\ref{Long}~and~\ref{Mult} we describe different
approaches to accommodate within-subject dependence.

\subsection{Marginal quantile model}\label{Marginal}
Denote $Y_{ij}$ as the measurement of SBP on individual $i = 1,2,\ldots,N$
at visit $j = 1,2,\ldots,J$ indexing the years 1991, 1993$,\ldots,$2009.
While in general $J$ can vary by individual, in our application $J$~is
constant across subjects.
This section describes a model for SBP that allows urbanization effects
to change along the distribution, with the extension to SBP and DBP in
Section~\ref{Mult}.
Let $\mathbf{X}_{ij}$ be a covariate vector of length $P$ containing
the variables such as age and urbanization for individual $i$ at visit $j$.
Denote the conditional distribution function of $Y_{ij}$ as $F(y|\mathbf{X}_{ij}) = P(Y_{ij}\leq y| \mathbf{X}_{ij})$.
We specify the distribution of absolutely continuous $Y_{ij}$ via its
quantile function, defined as $Q(\tau| \mathbf{X}_{ij}) = F^{-1}(\tau|
\mathbf{X}_{ij})$, where $\tau\in(0,1)$ is known as the quantile level.
For each response $Y_{ij}$ there exists a latent $U_{ij} \stackrel
{}{\sim} U(0,1)$ such that $Y_{ij} = Q(U_{ij} | \mathbf{X}_{ij})$.

We assume the quantile function of SBP $Q(\tau| \mathbf{X})$ is a
linear combination of covariates, that is,
\[
Q(\tau| \mathbf{X}) = \sum_{p = 1} ^{P} X_{p} \beta_{p}(\tau).
\]
The regression parameter $\beta_{p}(\tau)$ is the effect of the $p$th
covariate on $Q(\tau| \mathbf{X})$.
A~one-unit increase in $X_{p}$ is associated with a $\beta(\tau)$
increase in the $\tau$th population quantile.
We refer to $\beta(\tau)$ as a ``fixed effect,'' since this effect
applies to the full population.

Similar to \citet{RS2013}, we project $\beta_{p}$ onto a space of $M
\geq2$ parametric basis functions $I_{1}(\tau),\ldots, I_{M}(\tau)$
defined by a sequence of knots $0 = \kappa_{0} < \kappa_{1} <\cdots <
\kappa_{M} < \kappa_{M + 1} = 1$.
Let $q_{0}(\tau)$ be the quantile function of a random variable from a
parametric location/scale family with location parameter 0 and scale
parameter 1.
The basis functions are defined as $I_{1}(\tau) \equiv1$, $I_{2}(\tau)
= q_{0}(\tau)\mathbh{1}_{\tau\leq\kappa_{1}} + q_{0}(\kappa
_{1})\mathbh{1}_{\tau> \kappa_{1}}$, and
\[
I_{m}(\tau)= \cases{ 0, &\quad$\tau\leq\kappa_{m - 1}$,
\cr
q_{0}(\tau) - q_{0}(\kappa_{m - 1}), &\quad$
\kappa_{m - 1} \leq\tau\leq\kappa_{m}$,
\cr
q_{0}(
\kappa_{m}) - q_{0}(\kappa_{m - 1}), &\quad$\tau>
\kappa_{m}$}
\]
for $m > 2$.
Our model is of the form $\beta_{p}(\tau) = \sum_{m = 1}^{M} I_{m}(\tau
) \theta_{mp}$ and, thus,
\begin{eqnarray}
Q(\tau|X_{ij}) &=& \sum_{p = 1} ^{P}
X_{ijp} \beta_{p}(\tau) = \sum_{p = 1}
^{P} X_{ijp}\sum_{m = 1}^{M}I_{m}(
\tau)\theta_{mp}, \label{eq:QR_mod}
\end{eqnarray}
where $\theta_{mp}$ are the regression weights.
By partitioning our distribution by the knots, we only assume our
distribution is locally parametric, so our method is semiparametric.
We set X$_{ij1} \equiv1$ for all $i$ and $j$ for the intercept.

\begin{figure}

\includegraphics{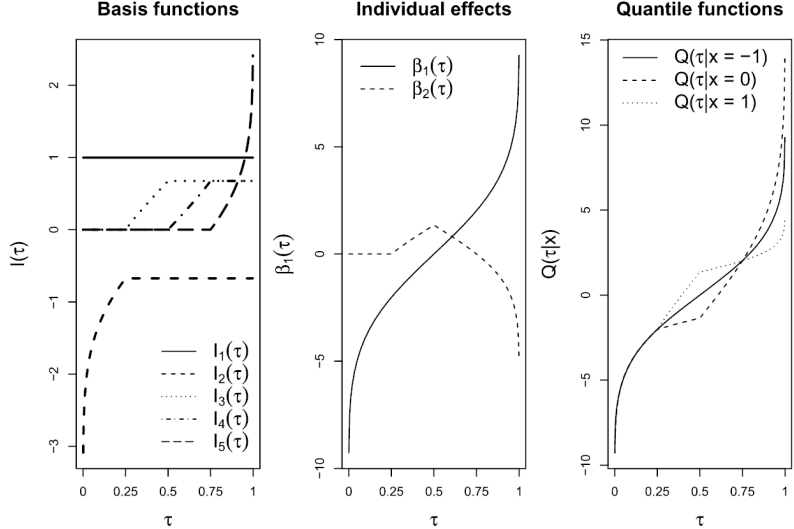}

\caption{Plots of the quantile model with $M = 5$ basis functions with
knots at (0.25, 0.5, 0.75) and one covariate $x$, with $\bolds
{\theta}_{.1} = (0,3,3,3,3)$ corresponding to basis coefficients for
the intercept and $\bolds{\theta}_{.2} = (0,0,2,-2,-2)$
corresponding to basis coefficients for $x$. The left plot displays the
constant basis function ($m = 1$) and 4 Gaussian basis functions that
each correspond to a different quartile of the distribution. The middle
plot displays the intercept process $\beta_{1}(\tau)$, which is the
distribution when all covariates are 0, and the covariate effect $\beta
_{2}(\tau)$, defined as the deviation in the intercept process due to a
one unit change in $x$. The final panel displays the conditional
quantile function $Q(\tau|x) = \beta_{1}(\tau) + x \beta_{2}(\tau)$ for
$x = -1$, $x = 0$, and $x = 1$.}
\label{f:basisfunctions}
\end{figure}

An example of our model is displayed in Figure~\ref{f:basisfunctions}.
The left panel shows an example of Gaussian basis functions, where
$q_{0}(\tau) = \Phi^{-1}(\tau)$, $\Phi(z)$ is the distribution function
of the standard normal distribution, with knots at $(0.25, 0.5, 0.75)$.
Only one basis function changes at each quantile of the distribution.
The middle panel illustrates the projection of these basis functions
for $\bolds{\theta}_{.1} = (0,3,3,3,3)$ corresponding to basis
coefficients for the intercept and $\bolds{\theta}_{.2} =
(0,0,2,-2,-2)$ corresponding to basis coefficients for lone covariate $x$
(e.g., urbanization score). The middle panel shows the quantile
function when $x = 0$ $(\beta_{1}(\tau))$ and how the covariate effects
on the quantile function change across the distribution $(\beta_{2}(\tau
))$. The final panel displays the conditional quantile function, which
is $(\beta_{1}(\tau)) + x (\beta_{2}(\tau))$, for $x \in(-1, 0, 1)$.
The effect of $x$ is 0 in the first quartile, positive in the second
quartile, and negative in the third and fourth quartiles.
%

To achieve a valid quantile function (i.e., increasing in $\tau$), we
map all predictors into the interval $[-1,1]$, and constrain the
regression parameters such that\ $\theta_{m1} > \sum_{p=2}^{P} |\theta
_{mp}|$ for $ m > 1$.
We model $\theta_{mp}$ as a function of a Gaussian random variable~$\theta^{\star}_{mp}$.
The regression parameter $\theta_{mp}$ is set to $\theta^{\star}_{mp}$
if the constraint is satisfied and set to
\[
\theta_{mp}= \cases{ 0.001, &\quad$p = 1$,
\cr
0, &\quad otherwise,}
\]
if $\theta^{\star}$ is outside of the constraint space.
Details are outlined in \citet{RS2013}.
The latent regression variables $\theta^{\star}_{mp}$ are given
Gaussian priors with means $\mu_{mp}$ and precisions $\iota^{2}_{mp}$.

Let $\bolds{\theta}_{m.}$ be the collection of regression
parameters associated with basis function $m$.
When the base quantile function is Gaussian [i.e., $q_{0}(\tau) = \Phi
^{-1}(\tau)$], if $M = 2$ or $\bolds{\theta}_{2.} =\cdots =
\bolds{\theta}_{M.}$, this model simplifies to a Gaussian
heteroskedastic regression model, where $Q(\tau) = \mathbf{X}_{ij}'\bolds{\theta}_{m1} + \mathbf{X}_{ij}'\bolds{\theta
}_{m2}\Phi^{-1}(\tau)$
and, thus, $Y_{ij}|\mathbf{X}_{ij} \sim N(\mathbf{X}_{ij}'\bolds{\theta}_{m1},(\mathbf{X}_{ij}'\bolds{\theta
}_{m2})^{2})$.
Standard Gaussian linear regression is a special case of the
heteroskedastic regression model where $M = 2$ and $\theta_{mp} \equiv
0$ for $m > 1$ and $p > 1$ and $Y_{ij}|\mathbf{X}_{ij} \stackrel{}{\sim
} N(\mathbf{X}_{ij}'\bolds{\theta}_{m1},\theta_{12}^{2})$.

\subsection{Mixed effects quantile model}\label{Long}
In this section we introduce a semiparametric model that extends the
standard Gaussian mixed effects model to the quantile regression domain.
This enables us to examine the effects of urbanization across the full
distribution while accounting for the longitudinal structure of CHNS
data through random effects.
We utilize Gaussian basis functions [$q_{0}(\tau) = \Phi^{-1}(\tau)$]
for connections to standard mixed models.
Recall the canonical Gaussian random effects model [\citet{Fitz2012}]
\begin{eqnarray}
\mathbf{Y}_{i} &=& \mathbf{X}_{i}\bolds{\beta} +
\mathbf{Z}_{i}\bolds{\gamma}_{i} +
\mathbf{E}_{i}, \label{eq:Gaussian_mixed_effects}
\end{eqnarray}
where $\bolds{\beta}$ is a vector of fixed effects,
$\mathbf{Z}_{i}$ is a $J$ by $R$ matrix of random effect covariates,
$\bolds{\gamma}_{i}\stackrel{\mathrm{i.i.d.}}{\sim}N(\mathbf{0},
\bolds{\Delta})$ is a vector of length $R$ of random effects
specific to unit $i$,
and $\mathbf{E}_{i}\stackrel{\mathrm{i.i.d.}}{\sim}N(\mathbf{0},
\bolds{\Lambda})$ are random errors.

We can rewrite (\ref{eq:Gaussian_mixed_effects}) in three forms.
Conditional on the random effects,
$\mathbf{Y}_{i}|\mathbf{X}_{i}, \mathbf{Z}_{i}\sim N(\mathbf{X}_{i}\bolds{\beta} + \mathbf{Z}_{i}\bolds{\gamma}_{i},
\sigma^{2}\mathbf{I})$.
Marginally over the random effects,
$\mathbf{Y}_{i}|\mathbf{X}_{i} \sim N(\mathbf{X}_{i}\bolds{\beta}, \bolds{\Psi}_{i})$, where
$\bolds{\Psi}_{i} = \mathbf{Z}_{i}\bolds{\Delta}\mathbf{Z}_{i}' + \bolds{\Lambda}$.
Finally, the marginal quantile function form is $Q(\tau|\mathbf{X}_{ij}) = \mathbf{X}_{ij}'\bolds{\beta} + \psi_{ij}\Phi^{-1}(\tau)$,
where $\psi_{ij}$ is the $j$th diagonal element of $\bolds{\Psi}_{i}$.
Therefore, $Y_{ij}|\mathbf{X}_{ij} = \mathbf{X}_{ij}'\beta+ \psi
_{ij}\Phi^{-1}(U_{ij})$, where $U_{ij} \sim   U(0,1)$
marginally, with dependence between the $U_{ij}$ within the same subject.

We use the third representation to extend mixed models to the quantile
domain by viewing the transformed response as a realization from a
potentially correlated Gaussian process.
To account for the within-subject dependence, we hierarchically model
the latent $U_{ij}$ through a Gaussian copula.
Our model is
\begin{eqnarray}\label{eq:Mixed_quantile_model}
Y_{ij} &=& \sum_{p = 1}^{P}X_{ijp}
\beta_{p}(U_{ij}),\nonumber
\\
U_{ij} &=& \Phi(W_{ij} / \sqrt{\psi_{ij}}),
\\
\mathbf{W}_{i} &=& \mathbf{Z}_{i}'\bolds{
\gamma}_{i} + \mathbf{E}_{i}. \nonumber
\end{eqnarray}
%
The fixed regression effects $\beta(\tau)$ are modeled as in Section~\ref{Marginal}.
As in (\ref{eq:Gaussian_mixed_effects}), the random effects $\bolds
{\gamma}_{i}\stackrel{\mathrm{i.i.d.}}{\sim}N(\mathbf{0}, \bolds
{\Delta})$ and random errors $\mathbf{E}_{i}\stackrel{\mathrm
{i.i.d.}}{\sim}N(\mathbf{0}, \bolds{\Lambda})$.

The copula in (\ref{eq:Mixed_quantile_model}) permits structured
dependence in the $U_{ij}$.
This preserves the interpretability of population-level quantile
effects $\beta_{p}$ and accounts for within-subject dependence,
enabling simultaneous inference at the population and subject levels.
This formulation allows predictors to have a complex relationship with
the response.
A covariate can have a different effect in the middle of the
distribution relative to the tails.
This is represented by the fixed component $\mathbf{X}'\bolds{\beta
}(\tau)$, the conditional $\tau$th population quantile, with the same
interpretation of covariate effects as in Section~\ref{Marginal}.
Further, individuals in a population are allowed to respond differently
to the same covariate.
This is represented by the random component $\mathbf{Z}_{i}'\bolds
{\gamma}$.
A~one unit increase in $Z_{ijr}$ is associated with a $\gamma_{ir} /
\sqrt{\psi_{ij}}$ increase in the $Z$-score of individual $Y_{ij}$.

For the CHNS data we anticipate that between-individual variability is
strong, which can be estimated through a random intercept inside the copula.
Covariates that change across time (e.g., urbanicity) can be used to
further capture within-subject variability.
For longitudinal data we anticipate serial within-subject correlation,
so we model $\bolds{\Lambda} = \mathbf{I} + \lambda\bolds{\Xi
}(\alpha)$ as the sum of an identity matrix and a scaled (by positive
$\lambda$) autoregressive order-1 (AR-1) correlation matrix $\bolds
{\Xi}(\alpha)$, where $\bolds{\Xi}(\alpha)[u,v] = \alpha^{ -|u -
v|}$ with correlation parameter $\alpha$.
The scaling factor $\lambda$ determines the proportion of variability
determined by the temporal signal.

While we have thus far defined our model in terms of Gaussian basis
functions, any of the parametric bases described in \citet{RS2013} can
be utilized to model effects at the population level.
Finally, the standard Gaussian mixed model is a special case of (\ref
{eq:Mixed_quantile_model}), where $q_{0}(\tau) = \Phi^{-1}(\tau)$, $M =
2$ and $\theta_{mp} \equiv0$ for $m > 1$ and $p > 1$.
This allows us to center our flexible model on the popular model.

We assume the distribution of SBP can be partitioned into $M - 1$
components such that within each partition the distribution of SBP
behaves parametrically.
This partitioning enables the error distribution to adapt across
components. For example, the error term can be different in the lower
tail, the middle of the distribution and the upper tail. Our error
distribution is dependent on the covariates, so even in the least
flexible fit $(M = 2)$ our model permits more flexibility than standard
mixed models.

We assume the within-subject dependence can be characterized through a
multivariate normal distribution for several reasons. Using Gaussian
basis functions and a Gaussian copula enables us to center our prior
distribution on the canonical Gaussian mixed model. The Gaussian copula
allows us to account for within-subject serial correlation, a potential
issue with CHNS data, and correlation structures that are affected by
covariates (e.g., urbanicity). The Gaussian copula easily imputes
missing values, which was necessary for our application and its high
rate of missing values. Finally, the Gaussian copula is computationally
cheap. However, the price of the Gaussian copula is a lack of
flexibility, as we discuss in Section~\ref{Conclusions}. Nonparametric
copulas [\citet{Fuentes2013}] could be a useful extension in other applications.

\subsection{Multivariate mixed effects quantile model}\label{Mult}
Here we extend (\ref{eq:Mixed_quantile_model}) to the multivariate domain.
We are not concerned with trying to define a multivariate quantile
[\citet{Chak2003}], which imposes order on a collection of objects of
multivariate dimension.
The most common example of a multivariate quantile is a multivariate
median, which is a common alternative to the multivariate mean for
defining the center of a multivariate distribution.
Instead, we want to conduct simultaneous inference on observations with
multiple responses.
We anticipate SBP and DBP may have similar distributions, so by jointly
modeling them we can borrow information across responses.
Further, SBP and DBP are correlated within an individual, so we must
account for this dependence.
Otherwise, our estimates of uncertainty will be too small.
In summary, we are interested in conducting simultaneous inference at
the marginal medians (and other quantile levels) of SBP and DBP, rather
than defining a central point for both distributions.
A forerunner of our approach is \citet{Gilchrist2000}, who constructed
ordered quantile surfaces by reducing the dimension of a multivariate
random variable to one.
Our multivariate quantile regression approach allows us to
simultaneously analyze both responses and include covariates.
This allows us to draw inferences about how urbanization and other
covariates affect SBP and DBP, while preserving flexibility in how
changes in urbanization affect these distributions.
The approach in \citet{Gilchrist2000} does not permit covariates or
preserve the marginal distributions.

Denote $Y_{1ij}$ and $Y_{2ij}$ as the measurements of SBP and DBP on
individual $i$ at time $j$.
We specify different quantile effects for each response [i.e., $\beta
_{1p}(\tau_{1})$ and $\beta_{2p}(\tau_{2})$ for covariate $p$].
Our bivariate model then accounts for dependence between the parameters
in these quantile processes, and in the residual copula model.

Our multivariate mixed quantile model is
\begin{eqnarray} \label{eq:MV_mod}
Y_{hij} &=& \sum_{p = 1} ^{P}
X_{ijp}\sum_{m = 1}^{M}I_{m}(U_{hij})
\theta_{hmp},\nonumber
\\
U_{hij} &=& \bolds{\Phi}(W_{hij} / \sqrt{
\psi_{hij}}),
\\
\mathbf{W}_{i} &=& \mathbf{Z}_{i}\bolds{
\gamma}_{i} + \mathbf{E}_{i},\nonumber
\end{eqnarray}
where now $h = 1,2$ indexes the response of dimension $H$, $\mathbf
{W}_{i}$ is of length $JH$,
and the covariance of $\mathbf{E}_{i}$ is $\bolds{\Xi}(\alpha)
\otimes\bolds{\Lambda} + \mathbf{I}$ where $\bolds{\Xi
}(\alpha)[u,v] = \alpha^{ -|u - v|}$ with correlation parameter $\alpha
$ and $\bolds{\Lambda}$ is an unstructured $H \times H$
correlation matrix.

This formulation allows the uniform random variables $U_{ij}$ to be
interpreted as the individual's percentile relative to the population.
That is, an individual may be at the conditional $70$th percentile
($U_{1ij} = 0.70$) for SBP and the $75$th percentile ($U_{2ij} = 0.75$)
for DBP, and the similarity in these percentiles can be exploited.

To borrow strength across the responses, we model
$(\theta_{1mp}, \theta_{2mp})' \sim\break  \mathit{BVN}(\mu_{mp}\mathbf
{1}, \iota^{2}_{mp}\mathbf{I}_{2})$.
By shrinking regression effects to a common location, we are able to
borrow information across SBP/DBP to estimate covariate effects.
This multivariate framework enables statements about joint effects of a
predictor, and allows for probabilistic estimates regarding both
responses (e.g., the conditional probability an individual has blood
pressure higher than $140/90$).

We assign $\mu_{mp}$ independent normal priors with mean $\mu_{0mp}$
and precision $\iota^{2}_{0mp}$.
We give $\iota_{mp}$ independent Gamma$(a_{mp}, b_{mp})$ priors.
We designate $\bolds{\Lambda}$ an inverse Wishart prior with scale
matrix $\bolds{\Lambda}_{0}$ and $\nu_{0}$ degrees of freedom.
For our application we assign $\bolds{\Delta}$ a diagonal matrix
structure with diagonal elements $\delta_{hr} \stackrel{\mathrm
{i.i.d.}}{\sim} \operatorname{Gamma}(1,1)$, $h = 1,2,\ldots,H, r = 1,2,\ldots,R$.
In applications with more observations per subject and with correlation
on the within-subject regression coefficients easier to detect, more
complicated structures for $\bolds{\Delta}$ could be useful.
We use the Metropolis within Gibbs algorithm to sample from the
posterior, with details in the supplementary material [\citet{Smith2015}].

\section{Simulation study}\label{Sim}
We conducted a simulation study to examine the effect of within-subject
dependence on parameter estimation.
To construct univariate, auto-correlated responses, we generated dependent
$J$-dimensional \mbox{realizations} $W_{i} \sim N(0,\bolds
{\Psi_{i}})$, where $\bolds{\Psi}_{i} = \mathbf{Z}_{i}\bolds
{\Delta}\mathbf{Z}_{i}' + \bolds{\Xi}(\alpha) + \mathbf{I} $ with
$j$th diagonal element~$\psi_{ij}$.
The design matrix $\mathbf{Z}_{i}$ contains an intercept and one
continuous predictor $X_{1ij} \stackrel{\mathrm{i.i.d.}}{\sim} U(-1,1)$.

The first factor we examine in the simulation study is the strength of
the within-subject dependence.
We look at three levels, 0.0, 0.5, 0.9, of the temporal correlation
parameter $\alpha$, which correspond to no, moderate and strong
within-subject dependence, respectively.
Our second factor is the strength of the dependence determined by the
covariance of the within-subject random effects, $\bolds{\Delta} =
\Delta\mathbf{I}_{2}$.
In one setting the variance $\Delta= 0$, corresponding to the
coefficients having no effect on the dependence.
In the other $\bolds{\Delta}$ is a diagonal matrix with nonzero
values of $\Delta= 3$, corresponding to roughly $60\%$ of the variance
within a subject being explained by covariates.

Given these correlated responses, we perform the probability integral
transform $U_{ij} = \Phi(W_{ij}/ \sqrt{\psi_{ij}})$.
The third factor in our study is the marginal distribution given these
uniform random variables.
The response data are
\begin{eqnarray*}
(1)&\qquad& Y_{ij} = 3\Phi^{-1}(U_{ij}) +
(X_{1ij} + X_{2ij}) (0.5 - U_{ij}) *
\mathbh{1}_{U_{ij} < 0.5},
\\
(2)&\qquad& Y_{ij} = (3 + X_{1ij} + X_{2ij})Q_{t}(U_{ij}),
\end{eqnarray*}
where $Q_{t}$ is the quantile function of Student's $t$-distribution
with 5 degrees of freedom,
$i = 1,2,\ldots,N$ individuals, and $j = 1,2,\ldots,J$ visits.
The covariate $X_{2i}$ is binary with equal probability of $-$1 and 1,
and is constant over time.
Design~(2) is a heteroskedastic linear model, but design (1) is more
challenging to fit, with nonzero effects for only half of the distribution.
We generated data at $J = 7$ timepoints for $N = 50$ and $N = 100$ individuals.
For each level of our design we ran 100 Monte Carlo replications.

We examine three competitors for our simulation study.
The first is the marginal quantile model of Section~\ref{Marginal}.
This model assumes independent replications within an individual.
The second model is the mixed effects quantile model of Section~\ref{Long}.
This model can account for serial correlation and subject-specific effects.
For both of these two models we fit 2, 3 and 5 basis functions for two
different parametric bases (Gaussian and Student's $t$).
For each Monte Carlo replication the final model is selected by having
the highest log psuedo marginal likelihood [\citet{Ibrahim2005}] across
number of basis functions and parametric bases.
For data type (1), the log pseudo marginal likelihood (LPML) most
commonly selected 5 Gaussian basis functions for the independent model
without a copula and 5 Student's $t$-distributed basis functions for
the copula model.
For data type~(2), the LPML most commonly selected 2 Student's
$t$-distribution basis functions for both models.
The third competitor is the model of \citet{RBW2010} using 25
approximation terms, denoted ``RBW.''
RBW is able to fit marginal effects while accounting for a random
intercept, but ignores temporal correlation covariate effects within a subject.

Prior means for $\theta_{mp}^{\star}$ were 1 for $p = 1$ and 0
otherwise, and prior variances were~10.
For the copula model we set (scalar) $\Lambda_{0} = 1$ and $\nu_{0} =
3$, corresponding to a prior mean of 4 and infinite variance for
$\bolds{\Lambda}$.
For the Student's $t$-distribution basis functions we gave the shape
parameter a normal prior on the log scale with mean $\log(10)$ and
variance $\log(10) / 2$.
Averages of coverage probability (CP) of 95\% intervals and mean
squared error (MSE) of each model evaluated at the quantile levels
$\tau=0.1,0.3,0.5,0.7,0.9$ for the $N = 50$ case are shown in Table~\ref{t:N_50}.
All of the conclusions listed below similarly held for the $N = 100$
case, shown in the supplementary material
[\citet{Smith2015}].

%
\begin{table}
\tabcolsep=0pt
\caption{Coverage probability (CP) and mean squared error (MSE) for
the $N = 50$ arm of the simulation study.
Nominal coverage probability is $95\%$.
We compare treating the data as independent within a subject
(``Ind''), fitting with a~copula (``Cop''), and the random effects
model of \citet{RBW2010} (``RBW'').
Coverage and MSE were evaluated at and averaged over the quantile
levels $ \{0.1,0.3,0.5,0.7,0.9 \}$.
For datatype${} = 1$, MSE values are less than depicted values by a factor
of 10. Estimators whose MSE were statistically significantly different than
the copula model are indicated by $^{*}$}\label{t:N_50}
\begin{tabular*}{\tablewidth}{@{\extracolsep{\fill}}@{}lcccccccccccc@{}}
\hline
& \multicolumn{6}{c}{\textbf{Coverage}} & \multicolumn{6}{c@{}}{\textbf{MSE}}\\[-6pt]
& \multicolumn{6}{c}{\hrulefill} & \multicolumn{6}{c@{}}{\hrulefill}\\
& \multicolumn{2}{c}{$\bolds{\alpha = 0.0}$} & \multicolumn{2}{c}{$\bolds{\alpha =0.5}$} & \multicolumn{2}{c}{$\bolds{\alpha = 0.9}$}
& \multicolumn{2}{c}{$\bolds{\alpha = 0.0}$} & \multicolumn{2}{c}{$\bolds{\alpha =0.5}$} & \multicolumn{2}{c@{}}{$\bolds{\alpha = 0.9}$}\\[-6pt]
& \multicolumn{2}{c}{\hrulefill} & \multicolumn{2}{c}{\hrulefill} & \multicolumn{2}{c}{\hrulefill}
& \multicolumn{2}{c}{\hrulefill} & \multicolumn{2}{c}{\hrulefill} & \multicolumn{2}{c@{}}{\hrulefill}\\
& \textbf{X1} & \textbf{X2} & \textbf{X1} & \textbf{X2} & \textbf{X1} & \textbf{X2} & \textbf{X1} & \textbf{X2} & \textbf{X1} & \textbf{X2} & \textbf{X1} & \textbf{X2} \\
\hline
& \multicolumn{12}{c@{}}{$\Delta= 0$, Datatype${}= 1$}\\
{Ind} & 0.91 & 0.95 & 0.88 & 0.94 & 0.73 & 0.94 & 0.05\phantom{$^{*}$} & 0.09\phantom{$^{*}$} & 0.07\phantom{$^{*}$} & 0.10\phantom{$^{*}$} & 0.11\phantom{$^{*}$} & 0.11\phantom{$^{*}$} \\
{Cop} & 0.95 & 0.96 & 0.95 & 0.97 & 0.94 & 0.96 & 0.05\phantom{$^{*}$} & 0.10\phantom{$^{*}$} & 0.06\phantom{$^{*}$} & 0.10\phantom{$^{*}$} & 0.09\phantom{$^{*}$} & 0.10\phantom{$^{*}$} \\
{RBW} & 0.83 & 0.90 & 0.86 & 0.89 & 0.88 & 0.87 & $0.11^{*}$ & $0.16^{*}$ & $0.13^{*}$ & $0.15^{*}$ & $0.16^{*}$ & $0.14^{*}$
\\[3pt]
& \multicolumn{12}{c@{}}{$\Delta= 0$, Datatype${} = 2$}\\
{Ind} & 0.94 & 0.97 & 0.89 & 0.94 & 0.76 & 0.95 & 0.06\phantom{$^{*}$} & 0.09\phantom{$^{*}$} & 0.09\phantom{$^{*}$} & 0.11\phantom{$^{*}$} & 0.15\phantom{$^{*}$} & 0.13\phantom{$^{*}$} \\
{Cop} & 0.98 & 0.98 & 0.95 & 0.98 & 0.93 & 0.97 & 0.07\phantom{$^{*}$} & 0.11\phantom{$^{*}$} & 0.10\phantom{$^{*}$} & 0.13\phantom{$^{*}$} & 0.16\phantom{$^{*}$} & 0.13\phantom{$^{*}$} \\
{RBW} & 0.96 & 0.98 & 0.95 & 0.96 & 0.93 & 0.95 & 0.07\phantom{$^{*}$} & 0.10\phantom{$^{*}$} & 0.10\phantom{$^{*}$} & 0.11\phantom{$^{*}$} & 0.14\phantom{$^{*}$} & 0.10\phantom{$^{*}$}
\\[3pt]
& \multicolumn{12}{c@{}}{$\Delta= 3$, Datatype${} = 1$}\\
{Ind} & 0.61 & 0.76 & 0.58 & 0.78 & 0.56 & 0.72 & 0.17\phantom{$^{*}$} & 0.24\phantom{$^{*}$} & 0.19\phantom{$^{*}$} & $0.25^{*}$ & 0.23\phantom{$^{*}$} & $0.24^{*}$ \\
{Cop} & 0.92 & 0.91 & 0.91 & 0.90 & 0.89 & 0.91 & 0.14\phantom{$^{*}$} & 0.18\phantom{$^{*}$} & 0.15\phantom{$^{*}$} & 0.17\phantom{$^{*}$} & 0.18\phantom{$^{*}$} & 0.17\phantom{$^{*}$} \\
{RBW} & 0.85 & 0.70 & 0.85 & 0.69 & 0.86 & 0.67 & $0.23^{*}$ & $0.26^{*}$ & 0.24\phantom{$^{*}$} & $0.26^{*}$ & 0.25\phantom{$^{*}$} & $0.24^{*}$
\\[3pt]
& \multicolumn{12}{c@{}}{$\Delta= 3$, Datatype${} = 2$}\\
{Ind} & 0.64 & 0.76 & 0.60 & 0.78 & 0.57 & 0.74 & 0.26\phantom{$^{*}$} & 0.26\phantom{$^{*}$} & 0.28\phantom{$^{*}$} & 0.25\phantom{$^{*}$} & $0.32^{*}$ & $0.29^{*}$ \\
{Cop} & 0.90 & 0.90 & 0.89 & 0.89 & 0.91 & 0.92 & 0.21\phantom{$^{*}$} & 0.21\phantom{$^{*}$} & 0.20\phantom{$^{*}$} & 0.21\phantom{$^{*}$} & 0.19\phantom{$^{*}$} & 0.17\phantom{$^{*}$} \\
{RBW} & 0.86 & 0.80 & 0.84 & 0.79 & 0.83 & 0.77 & 0.26\phantom{$^{*}$} & 0.21\phantom{$^{*}$} & 0.26\phantom{$^{*}$} & 0.22\phantom{$^{*}$} & $0.31^{*}$ & 0.24\phantom{$^{*}$}\\
\hline
\end{tabular*}
\end{table}

When observations within a subject are independent ($\Delta= 0$, $\alpha
= 0$ case), all models attain the nominal $95\%$ coverage probability
for both predictors and data types,
except RBW for data type (1).
Fitting a copula to independent data seems to have little effect on
marginal inference.
Increasing $\alpha$ causes undercoverage in the independent model for
the continuous predictor X$_{1}$.
In contrast, the copula and RBW models maintain proper coverage.
As within-subject dependence increases, each observation contributes
less information about the marginal distribution.
This can be seen by the increases in MSE due to increases in $\alpha$.
We compare MSE across the estimators when the covariates do not affect
within-subject dependence (i.e., $\Delta= 0$).
The copula model is better than RBW with respect to MSE for data type (1).
RBW assumes the heteroskedastic model, as in data type (2), yet none of
the three models are statistically significantly better with respect to MSE.

The results change when the subject-level regression coefficients
affect dependence (i.e., $\Delta= 3$).
RBW and the independent model suffer from poor coverage when the
predictors account for dependence in the response.
In contrast, the copula model maintains close to nominal coverage.
Further, the copula model dominates RBW and the independent model with
respect to MSE.
The copula model has a statistically significantly lower MSE in roughly
half of the cases and is lowest in all cases.
In summary, accounting for covariates in the dependence can reduce MSE
and preserve coverage.

\section{CHNS analysis}\label{Analysis}
In this section we analyze the CHNS data.
Our final sample consisted of 1421 females missing 56\% of blood
pressure measurements and 1248 males missing 55\% of blood pressure
measurements.
Missing household income in year $j$ was imputed using the community
average for year $j$.
Missing smoking status was imputed using the value from the previous
sampling wave, and assumed to be a nonsmoker in the first wave if missing.
Missing pregnancy status was assumed to be not pregnant.

With a large number of predictors and so many missing observations,
allowing all 14 predictors to change with quantile level is not feasible.
In our analysis we have urbanicity change with quantile level and all
other effects be constant with quantile level, that is, we fix $\beta
_{p}(\tau) \equiv\beta_{p} = \theta_{1}$ for all $\tau$ by setting
\mbox{$\theta_{2} =\cdots = \theta_{m} = 0$}.
The interpretations for these effects are equivalent to those in mean
regression in that they are allowed to affect the location but not the
shape of the response distribution.

Another challenge presented by these data is confounding due to blood
pressure medication.
Medication artificially suppresses blood pressure values.
For individuals on medication we ignore the measured values and assume
only that they have high blood pressure.
Using the method of \citet{RS2013}, we treat these values as
right-censored above the thresholds for high blood pressure, located at
140 for SBP and 90 for DBP.
For individuals on blood pressure medication the censored likelihood is
$p_{1} = P(Y1_{i}|\mathbf{X}_{i} > 140) = 1 - F_{Y1}(140|\mathbf
{X}_{i})$
for SBP and
$p_{2} = P(Y2_{i}|\mathbf{X}_{i} > 90) = 1 - F_{Y2}(90|\mathbf
{X}_{i})$
for DBP.
We use these censored probabilities in the likelihood for these individuals.

We linearly transformed the responses to have mean 0 and standard
deviation 1.
We assigned $\mu_{m1} \stackrel{\mathrm{i.i.d.}}{\sim} N(1,1)$ priors
for the intercept process and $m > 1$ and $\mu_{mp} \stackrel{\mathrm
{i.i.d.}}{\sim} N(0,1)$ priors for all other regression parameters.
We set $\Lambda\sim IW(10, 7\bolds{\Lambda}_{0})$,
where $\bolds{\Lambda}_{0}$ is an $H \times H$ correlation matrix
with off-diagonal elements of 0.
This corresponds to a prior mean of 1 and variance of 0.4 for the
diagonal elements of $\Lambda$.
This centers the prior distributions of SBP and DBP on a mean zero,
unit variance normal distribution that is independent across SBP and DBP.
We assigned $\iota_{mp}^{2} \stackrel{\mathrm{i.i.d.}}{\sim} G(1,1)$ priors.
We assigned $\alpha$ a uniform prior on the unit interval.
We ran our models for 40,000 MCMC iterations, the first half of which
we discarded.

\subsection{Analysis}
We fit 3 different models to compare dependence structures.
In model 1 we fit our model without a copula, assuming independence
across sampling wave and response.
We also fit two copula models.
In model 2 the covariance of $\mathbf{W}_{i}$ is $\bolds{\Xi
}(\alpha) \otimes\bolds{\Lambda} + \mathbf{I}_{JH}$, where the
nondiagonal component is the Kronecker product of an AR-1 correlation
matrix and an unstructured $2 \times2$ covariance matrix $\bolds
{\Lambda}$.
In model 3 we fit a mean component consisting of a random intercept and
an urbanicity effect with the same covariance as model 2, that is,
$\mathbf{W}_{i} = \mathbf{Z}_{i}\bolds{\gamma}_{i} + \mathbf{E}_{i}$.
Finally, we fit SBP and DBP jointly and singly for all copula models.
For each model we fit $M = 2$ and $M = 4$ basis functions.
The runs with 4 basis functions had convergence issues, probably due to
the large number of missing observations, so we present results for the
$M = 2$ case.

\begin{figure}

\includegraphics{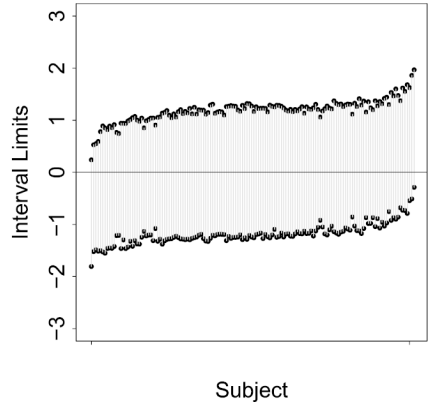}

\caption{Plots of posterior credible sets of urbanicity random effects
on females for systolic blood pressure.
For visual clarity posterior credible sets are ordered by posterior
median and every 10th subject is shown.}
\label{f:ci}
\end{figure}

LPML values were $-$32,060, $-$17,681 and $-$34,282 for females for models 1, 2
and 3 ($-$27,683, $-$15,576 and $-$29,310 for males).
The large values for model 2 indicate strong within-subject correlation
that is captured in the covariance.
The small LPML value for model 3 indicates that including
subject-specific slopes for urbanization leads to overfitting.
Figure~\ref{f:ci} illustrates the urbanicity random effect $\bolds
{\gamma}_{i2}$ on systolic blood pressure across female subjects.
These effects are not statistically significant.
Nonzero slope effects combined with missing observations can lead to
estimating many extreme quantile levels for the first and last sampling waves.
In applications with fewer missing observations or more timepoints,
random slope effects could be useful.

%

The posterior means of the off-diagonal elements of the correlation
between responses were 0.94 and 0.95 for females and males,
respectively, with posterior standard deviations around 0.01.
Therefore, SBP and DBP at one timepoint within an individual are very
strongly correlated, and the posterior distribution of the correlation
effects is dominated by the data.
The posterior means of the temporal correlation parameter $\alpha$ were
0.12 and 0.10 for females and males, respectively, with posterior
standard deviations around 0.02.
For the univariate fits of blood pressure the posterior means of $\alpha
$ ranged from 0.70 to 0.82 with posterior standard deviations around 0.02.
The multivariate and temporal correlation seem to be fighting for the
same signal.
This strong correlation within an individual across response and time
is useful when imputing missing values.

For females, the average of the posterior variance of the regression
effects at the median $\beta_{p}(0.5)$ was 3.59 for the multivariate
copula model and 4.24 for the multivariate independent model.
This 13\% increase in posterior variance (5\% for males) suggests the
independent model may be susceptible to undercoverage.
For females, the mean of the posterior variance of the regression
effects at the median was 3.62 for the univariate copula model.
This 2\% decrease in posterior variance for the multivariate model (1\%
for males) suggests that covariate effects are similar across SBP and DBP.
In applications where multivariate observations within an individual
were less correlated, we would expect a larger reduction in posterior
variance of the effects.
In summary, the copula models account for the within-subject dependence
and are less susceptible to undercoverage than models that assume
independent replications within an individual.
The multivariate quantile approach reduces posterior variance by
modeling SBP and DBP jointly.


%
\begin{figure}[b]

\includegraphics{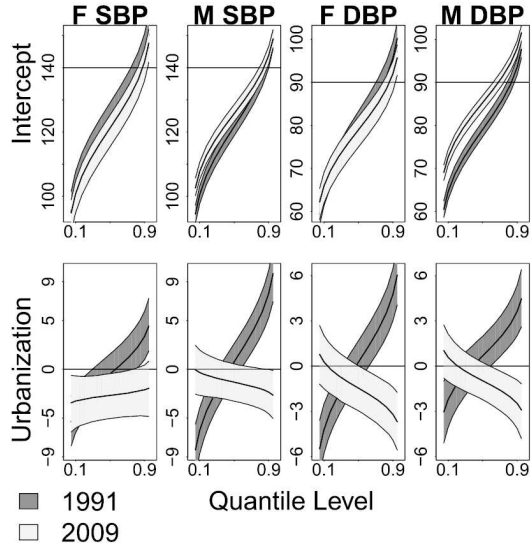}

\caption{Plots of the intercept process for the age 40--50 cohort and
population urbanicity effects by gender and blood pressure type.
The intercept process is the distribution of the response when all
covariates are 0.
The urbanicity plots are the effects of a one standard deviation
increase in urbanicity on the $\tau$th quantile of blood pressure.
Dark regions correspond to 1991 estimates, while light regions
correspond to 2009 estimates.}\vspace*{-3pt}\label{f:timeplots}
\end{figure}

Posterior plots of the intercept process $\beta_{1}(\tau)$ for the age
40--50 cohort and population urbanicity effects are shown in Figure~\ref{f:timeplots}.
The intercept process represents the values of our baseline age 40--50
cohort when all predictors are zero, which is a central value after
transformation to $[-1,1]$ for all covariates.
For the intercept process, the light 2009 regions differ statistically
from the dark 1991 regions for males in the lower tails of SBP and DBP.
In contrast to the intercept process, the urbanicity effects change
qualitatively from the first to the last sampling wave.
In 1991 urban areas had higher blood pressure in the upper tail and
lower blood pressure in the lower tail.
In 2009 the urbanicity effect is negative or zero for SBP for all
quantile levels.
In contrast, urban areas are now associated with lower DBP in the upper
tail of the distribution.
Figure~\ref{f:exploratory} illustrates that both blood pressure values
and urbanization have increased over time, while Figure~\ref{f:timeplots}\vadjust{\eject} shows that urban areas now have similar or lower
quantiles of SBP and DBP than their rural counterparts.
This indicates that rural areas are driving the increases in the upper
tails of the distributions of SBP and DBP.

%
\begin{sidewaystable}
\tabcolsep=0pt
\tablewidth=\textwidth
\caption{Posterior parameter estimates and 95\% credible intervals for location effects.
Mean effects include age cohort (with baseline group aged 40--50),
urbanicity by age cohort interaction (indicated by~$U*$),
household income (HHI), current pregnancy status and smoking}\label{t:means}
\begin{tabular*}{\tablewidth}{@{\extracolsep{\fill}}@{}ld{2.1}cd{2.1}cd{2.1}cd{2.1}c@{}}
\hline
\textbf{Predictor} & \multicolumn{2}{c}{\textbf{SBP 1991}} & \multicolumn{2}{c}{\textbf{SBP 2009}} & \multicolumn{2}{c}{\textbf{DBP 1991}} & \multicolumn{2}{c@{}}{\textbf{DBP 2009}}\\
\hline
\multicolumn{9}{c}{\textit{Female effects}}\\
18 $<$ Age $<$ 30 & -5.5 & $(-6.9,-3.9)$ & -4.2 & $(-5.9,-2.7)$ & -2.8 & $(-3.7,-2.0)$ & -3.2 & $(-4.3,-2.2)$ \\
30 $<$ Age $<$ 40 & -4.5 & $(-5.9,-3.3)$ & -1.5 & $(-2.8,-0.0)$ & -2.6 & $(-3.5,-1.8)$ & -0.2 & $(-1.1,0.7)$ \\
50 $<$ Age $<$ 60 & 2.7 & $(1.0,4.3)$ & 3.2 & $(1.8,4.6)$ & 1.3 & $(0.1,2.3)$ & 1.4 & $(0.5,2.3)$ \\
60 $<$ Age & 7.7 & $(6.0,9.6)$ & 5.5 & $(4.1,6.9)$ & 3.1 & $(2.0,4.4)$ & 1.9 & $(0.9,2.7)$ \\
$U * ($18 $<$ Age $<$ 30) & -0.9 & $(-5.7,3.6)$ & 1.1 & $(-4.1,7.1)$ & 1.2 & $(-2.0,4.2)$ & 3.6 & $(0.3,7.4)$ \\
$U * ($30 $<$ Age $<$ 40) & -4.5 & $(-10.1,1.5)$ & 0.8 & $(-4.1,6.2)$ & -0.8 & $(-4.4,3.1)$ & -0.0 & $(-3.0,3.1)$ \\
$U * ($50 $<$ Age $<$ 60) & 4.2 & $(-2.8,10.9)$ & 0.4 & $(-4.9,5.5)$ & 2.2 & $(-2.5,7.0)$ & 2.0 & $(-1.2,5.1)$ \\
$U * ($60 $<$ Age) & 2.7 & $(-3.2,8.2)$ & 6.6 & $(1.6,11.4)$ & 1.9 & $(-2.1,5.6)$ & 1.3 & $(-1.7,4.7)$ \\
HHI & 4.0 & $(-0.7,8.2)$ & -2.0 & $(-3.6,-0.1)$ & 2.2 & $(-0.6,4.8)$ & -0.8 & $(-1.9,0.4)$ \\
Pregnant & -1.1 & $(-4.5,2.9)$ & 2.0 & $(-1.8,4.8)$ & -1.0 & $(-2.8,1.6)$ & -0.3 & $(-3.2,1.9)$
\\[3pt]
\multicolumn{9}{c}{\textit{Male effects}}\\
18 $<$ Age $<$ 30 & -1.9 & $(-3.6,-0.5)$ & -4.0 & $(-5.7,-2.5)$ & -1.6 & $(-2.5,-0.5)$ & -3.3 & $(-4.2,-2.4)$ \\
30 $<$ Age $<$ 40 & -1.3 & $(-3.0,0.0)$ & -2.1 & $(-3.3,-0.7)$ & -1.0 & $(-2.0,0.1)$ & -1.2 & $(-2.1,-0.4)$ \\
50 $<$ Age $<$ 60 & 4.7 & $(3.4,6.5)$ & 0.8 & $(-0.4,2.1)$ & 2.1 & $(1.1,3.3)$ & 0.1 & $(-0.7,0.9)$ \\
60 $<$ Age & 8.3 & $(6.2,10.1)$ & 3.2 & $(1.9,4.5)$ & 3.6 & $(2.4,4.8)$ & 0.3 & $(-0.5,1.2)$ \\
$U * ($18 $<$ Age $<$ 30) & -0.2 & $(-5.8,4.7)$ & -0.5 & $(-5.8,4.9)$ & -1.6 & $(-5.2,1.8)$ & 3.1 & $(-0.4,6.6)$ \\
$U * ($30 $<$ Age $<$ 40) & -5.2 & $(-10.1,-0.6)$ & -1.2 & $(-6.1,3.5)$ & -4.0 & $(-7.6,-0.5)$ & 1.5 & $(-1.7,4.4)$ \\
$U * ($50 $<$ Age $<$ 60) & 5.7 & $(-0.1,11.5)$ & 0.6 & $(-4.5,6.4)$ & 0.9 & $(-2.9,5.3)$ & 0.9 & $(-2.6,4.4)$ \\
$U * ($60 $<$ Age) & 7.0 & $(1.6,13.6)$ & 2.4 & $(-2.5,7.4)$ & 1.0 & $(-2.9,5.1)$ & -0.0 & $(-3.2,3.1)$ \\
HHI & 2.6 & $(-0.5,5.3)$ & -1.8 & $(-3.4,-0.2)$ & 0.9 & $(-1.0,3.0)$ & -0.7 & $(-1.8,0.3)$ \\
Smoke & -0.1 & $(-0.9,0.9)$ & 0.5 & $(-0.2,1.4)$ & -0.1 & $(-0.7,0.5)$ & 0.0 & $(-0.5,0.5)$ \\
\hline
\end{tabular*}
\end{sidewaystable}

Estimated location effects are presented in Table~\ref{t:means}, where
several general associations are apparent.
Blood pressure increases with age, as expected a priori.
The interaction effects between urbanization and age represent the
differences in urbanization effects across different age groups.
Other than the cohort older than 60 years of age, 0 is in or very near
the limits of the credible sets of interaction effects for both males
and females for all years.
This indicates there is little evidence that the urbanization effect
changes much with age for individuals aged 60 and below, indicating
that rural Chinese youth may be more at risk for hypertension.
However, there are very few young individuals measured in later waves
(only 91 in 2006, 0~in 2009), so estimates for this cohort are less
stable and have larger variances.
For Chinese aged 60 and above the urbanization interaction effect is
positive for 2009.
This tends to bring the effect in the upper tails closer to 0, and
reduces the discrepancy in urbanization effect for older Chinese.
The covariates household income, pregnancy and smoking status have
little effect.
To examine the robustness of assuming a male was a nonsmoker if smoking
status was missing, we reran our final model assuming the individual
was a smoker instead of a nonsmoker if the first wave was missing. The
smoking effect was unchanged.

\section{Discussion}\label{Conclusions}
In this paper we have presented novel methods for analysis using mixed
models in a quantile regression framework to address a major limitation
in the hypertension literature: the inability to consider continuous
SBP and DBP as correlated outcomes.
Most hypertension literature either considers the discrete hypertension
outcome or continuous SBP or DBP outcomes in separate models.
Our quantile regression model enabled the exposure effect of
urbanization to vary smoothly along the distributions of SBP and DBP,
offering much more flexibility than mean regression models or models
that specify cutpoints.
We conducted a simulation study that illustrates the utility of
estimating dependence in SBP and DBP for quantile regresssion in a
longitudinal setting.
We found strong evidence of dependence of SBP and DBP and serial
correlation within an individual.
We found that the effects of the covariates are similar across SBP and
DBP, and inference can be enhanced by borrowing information across outcomes.
There are many biostatistical and epidemiological applications where
cutpoints are currently utilized to enable separate inference at
different parts of the distribution, including analyses of air
pollution, nutrients and apolipoprotiens, to name a few.
Our model obviates the choice of cutpoints.
This is a key advantage, as inference is often nonrobust to 
cutpoint selection.

We found that urbanicity is now associated with lower rather than
higher blood pressure, especially in the upper tails of the
distribution, potentially illuminating the segment of the population at
highest risk relative to dietary or physical activity changes occurring
with modernization.
It is possible that modernization-related changes of urbanization lead
to more protective lifestyle habits for individuals at the highest
levels of urbanization.
Perhaps these individuals have greatest access to health care and
environmental supports for healthy diet and physical activity, which
leads to some degree of protection at the upper tail of the distribution.
Our findings suggest that attention be paid to the center of the
urbanization distribution to address individuals who might be in
urbanizing areas, but without access to supports for healthy lifestyle
behaviors.
Given that urbanization has increased over the last twenty years and
the urbanization effect in the upper tail has diminished, blood
pressure appears to be increasing in China even in less urbanized areas.

We flexibly model the population level regression effects using a
linear combination of parametric basis functions.
We model the within-subject level dependence using a Gaussian copula.
We chose a Gaussian copula to facilitate centering of the prior
distribution, to accommodate within-subject serial correlation and
covariates affecting the within-subject dependence, and for simplicity
in computation and imputation of missing values.
However, these advantages may not be worth the restrictive behavior of
the Gaussian copula in other applications, and this is an area of
future research.
Gaussian copulas assume independence in the deep tails and assume the
same dependence in the lower tail as the upper tail.
In this paper we focus on the quantile levels from 0.1 to 0.9.
In practice, if inference at more extreme quantile levels is of
interest, other copulas should be considered.
Another useful extension is a fully nonparametric approach to the
quantile function.


\section*{Acknowledgments}
We thank the Associate Editor and two anonymous referees for their
helpful comments.

Penny Gordon-Larsen is grateful to the Carolina Population Center (R24
HD050924) for general support.

\begin{supplement}[id=suppA]
\stitle{MCMC details and additional simulation results\\}
\slink[doi]{10.1214/15-AOAS841SUPP} 
\sdatatype{.pdf}
\sfilename{aoas841\_supp.pdf}
\sdescription{Supplementary materials include a description of the
MCMC algorithm and an additional table of simulation study results.}
\end{supplement}

%

\printaddresses
\end{document}